\documentclass[10pt,preprint2,longnamesfirst]{aastex}
\received{2001 June~1}
\revised{2001 August~6}
\accepted{2001 August~13}
\journalid{563}{2001 December~10}
\paperid{54166}
\newcommand{\mjybm}{\mbox{mJy~beam${}^{-1}$}}

\shortauthors{LaRosa et al.}
\shorttitle{Parallel Nonthermal Filaments Near the Galactic Center}

\begin{document}

\title{A New System of Parallel Isolated Nonthermal Filaments Near
	the Galactic Center: Evidence for a Local Magnetic Field Gradient}

\author{T.~N.~LaRosa\altaffilmark{1}}
\affil{Department of Biological and Physical Sciences,
	Kennesaw State University, 1000 Chastain Rd., Kennesaw, GA
	30144}
\email{ted@avatar.kennesaw.edu}

\altaffiltext{1}{Navy-ASEE Summer Faculty Fellow, Naval Research 
	Laboratory}

\author{T.~Joseph~W.~Lazio, Namir E.~Kassim}
\affil{Code~7213, Remote Sensing Division, Naval Research
	Laboratory, Washington, DC 20375-5351}
\email{lazio@rsd.nrl.navy.mil}
\email{kassim@rsd.nrl.navy.mil}

\begin{abstract}
We report the discovery of a system of isolated nonthermal filaments
approximately 0.5\arcdeg\ northwest (75~pc in projection) of
\hbox{Sgr~A}.  Unlike other isolated nonthermal filaments which show
subfilamentation, braiding of subfilaments, and flaring at their ends,
these filaments are simple linear structures and more closely resemble
the parallel bundled filaments in the Galactic center radio arc.
However, the most unusual feature of these filaments is that the
20/90~cm spectral index uniformly decreases as a function of length,
in contrast to all other nonthermal filaments in the Galactic center.
This spectral gradient may not be due to simple particle aging but
could be explained by a curved electron energy spectrum embedded in a
diverging magnetic field.  If so, the scale of the magnetic gradient
is not consistent with a large scale magnetic field centered on
Sgr~A$^{\star}$ suggesting that this filament system is tracing a
local magnetic field.
\end{abstract}

\keywords {Galaxy: center --- radio continuum}

\section{Introduction}\label{sec:intro}

The isolated nonthermal filaments (hereafter NTFs) observed in the
Galactic center (hereafter GC) are unique to that region.  They are
characterized by extreme length to width ratios (from~10 to $>$100),
highly polarized emission (30--70\%), strong magnetic fields 
($\sim 1$~mG) aligned along their long axis, and nonthermal spectra
(e.g., the 20/90 cm spectral indices, $S \propto \nu^\alpha$, range
from $\alpha= -0.4$ to~$-0.6$, typically steepening above~5~GHz to
$\alpha\sim-1.5$), for a review see \cite{ms96}.  Subsequent work has
revealed that many isolated NTFs consist of subfilaments braided
around each other \citep[e.g.,][]{y-zwp97, laklg99}, and that the
surface brightness appears to be a maximum at the intersection of the
subfilaments \citep{lklh00}.  Lastly, \cite{lme99} and \cite{lklh00}
found that the spectral indices in several well studied filaments (the
southern and northern threads and the Sgr~C filament) are constant
with length.  The exception is the Snake filament which exhibits a
spectral gradient in the region surrounding its major \lq\lq
kink\rq\rq \citep{gnec95}.
 
The relationship between the NTF phenomenon and the parallel bundled
filaments in the GC Radio Arc is not clear.  The
filaments in the Radio Arc show little substructure, have nearly flat
spectra and can be observed at frequencies as high as 150~GHz
\citep{rsm00}.  Several theoretical models suggest that the Arc
filaments and the NTFs are tracing a large-scale magnetic field that
pervades the GC region \citep[e.g,][]{sm94}. At present, there is no
consensus interpretation of these structures and new models are under
development \citep[e.g.,][]{sl99,bl01,desl01}.

A recent wide-field image of the GC region at~90~cm \citep{lklh00}
revealed several extended sources well away from the GC itself.  One
of these, \objectname[]{G358.85$+$0.47}, which is 225~pc in projection
from \objectname[]{Sgr~A$^{\star}$}, was found to be the first NTF
that is parallel to the Galactic plane \citep{laklg99}.  This paper
reports a detailed study of \objectname[]{G359.85$+$0.39}, another
source discovered on the wide-field image.  It is a system of three
parallel NTFs located approximately 0.5\arcdeg\ ($\approx 75$~pc in
projection) northwest of \objectname[]{Sgr~A}.  In \S\ref{sec:observe}
we present 90, 20, and 6~cm observations of this source, made with the
Very Large Array (VLA) of the National Radio Astronomy Observatory.\footnote{%
The National Radio Astronomy Observatory is a facility of the National
Science Foundation, operated under a cooperative agreement with the
Associated Universities, Inc.%
}  In \S\ref{sec:theory} we describe alternate scenarios to explain the
morphology and spectral index gradient of the source; we present our
conclusions in \S\ref{sec:conclude}.

\section{Observations and Analysis}\label{sec:observe}

\subsection{Observations}\label{sec:observations}

\begin{figure}
\vspace*{-1cm}
\epsscale{0.9}
\rotatebox{-90}{\plotone{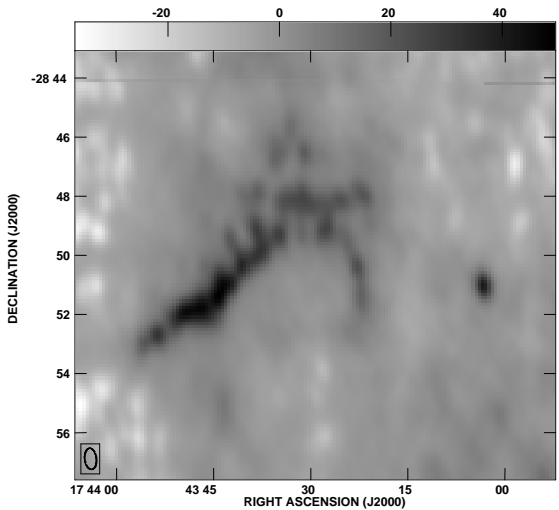}}
\caption[]{Grey scale representation of
\protect\objectname[]{G359.85$+$0.39} at~90~cm \citep{lklh00}.  The
resolution is $43\arcsec \times 23\arcsec$, the rms noise level is
5~\mjybm, and the gray scale is linear.}
\label{fig:90}
\end{figure}

The extended source \objectname[]{G359.85$+$0.39} was discovered
originally on a wide-field, 90~cm image of the GC region
\citep{lklh00}.  Figure~\ref{fig:90} is a subimage of the original
wide-field image and shows this source to consist of a linear
structure approximately 8\arcmin\ long that curves into a
semi-circular shape with a radius of~2\arcmin.  The surface brightness
over the linear region rises uniformly from about~20~\mjybm\ at the
southeastern end to a peak of~49~\mjybm\ near the center before
merging into the circular part in the northwest.  The surface
brightness in the circular region is somewhat patchy and ranges
from~15 to~25~\mjybm.  The rms noise on this image is about 5~\mjybm\
so we regard both the linear and semi-circular structures to be
significant.  At the distance of the GC (8.5~kpc) the 90~cm length
of~8\arcmin\ corresponds to a physical length of about 20 pc.  In
order to determine the spectrum and search for substructure in the
source, we obtained additional, higher-frequency observations.

\begin{figure}
\epsscale{1}
\plotone{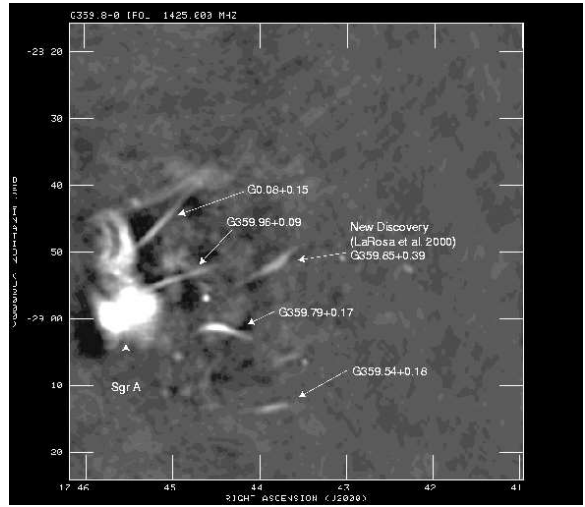}
\caption[]{Grey scale representation of the region surrounding
\protect\objectname[]{G359.85$+$0.39} at~20~cm.  The resolution is
$42.3\arcsec \times 38.5\arcsec$, the rms noise level is 2~\mjybm, and
the gray scale is linear.}
\label{fig:20}
\end{figure}

\begin{figure}
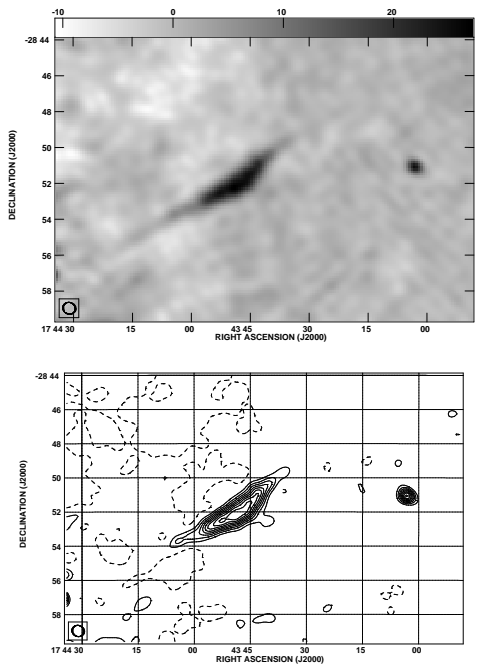

\begin{center}
\epsscale{0.6}
\rotatebox{-90}{\plotone{LaRosa_f3a.eps}}
\rotatebox{-90}{\plotone{LaRosa_f3b.eps}}
\end{center}
\vspace{-0.9cm}
\caption[]{\protect\objectname[]{G359.85$+$0.39} at~20~cm.  The
resolution is $42\farcs3 \times 38\farcs5$, and the rms noise level is
2~\mjybm.
\textit{Top}: The gray scale is linear.
\textit{Bottom}: The contour levels are 1~\mjybm\ $\times$ $-3$, 3, 6,
9, \ldots.}
\label{fig:20sub}
\end{figure}

We observed the source at~6 and~20~cm with the VLA;
Table~\ref{tab:log} summarizes the observations.  The 20~cm
observations were made in 1999 February with the VLA in the DnC
configuration.  This configuration provides a resolution comparable to
that of the 90~cm observations.  Standard processing techniques were
used to calibrate and image the visibility data.  Figure~\ref{fig:20}
shows the 20~cm DnC array image of the region centered on
\objectname[]{G359.85$+$0.39}.  This image shows both the northern and southern
threads, as well as the NTFs \objectname[]{G359.79$+$0.17} and
\objectname[]{G359.54$+$0.18}.  At this resolution
\objectname[]{G359.85$+$0.39} appears to be another isolated NTF with
a similar orientation as the other NTFs.  Figure~\ref{fig:20sub} shows
a subimage centered on \objectname[]{G359.85$+$0.39}.  The 20~cm
surface brightness is similar to the 90~cm brightness but peaks at a
different location.  The source is also approximately 50\arcsec\
shorter than at~90~cm, and the semi-circular emission is not detected.

\begin{deluxetable}{cccccc}
\tablewidth{0pc}
\tabletypesize{\small}
\tablecaption{VLA Observing Log\label{tab:log}}
\tablehead{
                      &                 & \colhead{VLA}           &
	 & \colhead{Integration} & \colhead{Synthesized} \\
 \colhead{Wavelength} & \colhead{Epoch} & \colhead{Configuration} &
	\colhead{Bandwidth} & \colhead{Time} & \colhead{Beam} \\
 \colhead{(cm)}       &                 &                         &
	\colhead{(MHz)} & \colhead{(hr)} & \colhead{(arcsecond)}}

\startdata
 6 & 2000 July~6      & DnC & 50 & 1.5 & $13 \times 9.5$ \\
20 & 1999 February~27 & DnC & 50 & 2.9 & $43 \times 38$ \\
\enddata
\tablecomments{We summarize only the new observations.  The 90~cm
observations are described in \cite{lklh00}.}
\end{deluxetable}

The semi-circular structure visible at~90~cm (Figure~\ref{fig:90}) is
notably absent at~20~cm (Figure~\ref{fig:20}).  Given the confused
nature of the GC, we cannot rule out conclusively the possibility that
this semi-circular emission is the superposition of another source,
but we regard it as unlikely.  First, its morphology is unlike that of
most extragalactic sources.  While its morphology is similar to that
of a (portion of a) supernova remnant, its spectrum is too steep.  Our
20~cm noise level is approximately 1~\mjybm, so a 3$\sigma$ upper
limit on its flux density implies a 20/90~cm spectral index steeper
than $-1.1$.  Second, the surface brightness from the linear part
merges smoothly with that of the semi-circular part at~90~cm.
Finally, the fact that the source is longer at 90~cm and that the
emission peaks at different locations at~20 and 90~cm suggests a
spectral index gradient.  Below we show that this source does indeed
have a spectral index gradient (from north to south) and the spectral
index at the north end of the the object is $-1.1$, consistent with
the spectral index of the semi-circular part being steeper than
$-1.1$.

In~2000 July, 6~cm continuum observations were made in the VLA DnC
configuration.  These observations were made in dual polarization mode
at~4515 and~4765 MHz.  Standard calibration and imaging techniques
were used to process the visibility data.  Because both the 6
and~20~cm observations were obtained in the DnC configuration, the
6~cm image has higher angular resolution.

\begin{figure}
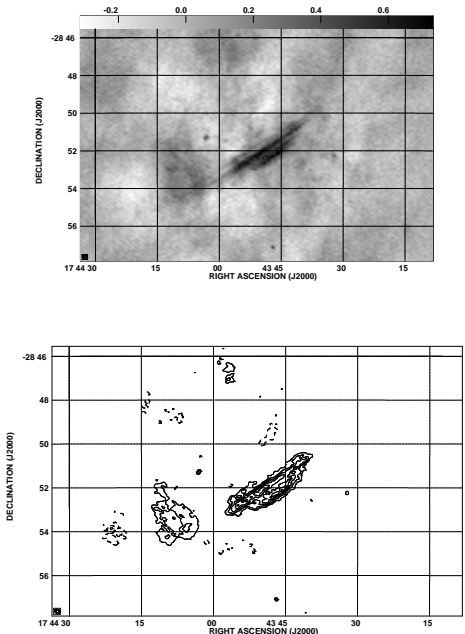

\begin{center}
\epsscale{0.6}
\rotatebox{-90}{\plotone{LaRosa_f4a.eps}}
\vspace{-0.25cm}
\rotatebox{-90}{\plotone{LaRosa_f4b.eps}}
\end{center}
\vspace{-0.75cm}
\caption[]{\protect\objectname[]{G359.85$+$0.39} at~6~cm.  The
resolution is $12\farcs5 \times 9\farcs3$, and the rms noise level is
0.08~\mjybm.
\textit{Top}: The gray scale is linear between~$-0.31$
and~0.73~\mjybm.
\textit{Bottom}: The contour levels are 0.1~\mjybm\ $\times$ $-0.225$, 
0.225, 0.325, 0.425, \ldots.}
\label{fig:6}
\end{figure}

Figure~\ref{fig:6} shows the resulting image.  The increased
resolution reveals that the linear structure consists of three
\emph{parallel} filaments.  As with the 20 and~90~cm emission the
surface brightness in all three filaments rises uniformly from the
southeastern end, peaks near the mid point, and declines toward the
northwestern end.  However, the ultra thin, top filament, although
detectable to the eye, is at best only a $3\sigma$ detection at some
points along its length.

\begin{figure}
\epsscale{0.8}
\rotatebox{-90}{\plotone{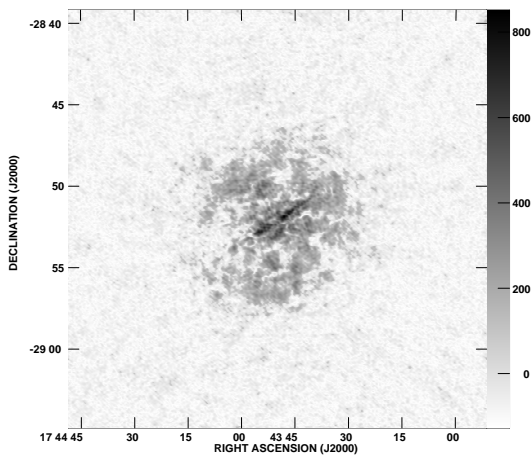}}
\caption[]{Total linear polarization of
\protect\objectname[]{G359.85$+$0.39} at~6~cm.  The resolution and
noise level are similar to that of Figure~\ref{fig:6}.  The polarized
emission from \protect\objectname[]{G359.85$+$0.39} is evident at the
center of the image, but there is considerable polarized flux that has
no correspondence in total intensity (see \S\ref{sec:observations})}
\label{fig:6pol}
\end{figure}

Figure~\ref{fig:6pol} shows the linear polarization image.  This image
indicates that filaments are strongly polarized but also shows
considerable polarized emission that does not correspond to any
features in total intensity.  A similar phenomonen has been found at
other wavelengths and in other directions \citep{gldt98,hkd00,gdmgwh00} and can
be explained by a foreground Faraday screen inducing polarization in a
diffuse background emission.  Analysis of the polarization is beyond
the scope of this work and will be discussed elsewhere.

We stress that the sub-filaments are parallel since that term has be
used to describe many isolated NTFs.  However, in all other isolated
NTF systems the filaments appear to overlap and cross
\citep{y-zwp97,lme99}.  Several also show flaring at their ends.  Even
at this fairly high resolution no substructure is detectable.  Thus,
from a morphological perspective these filaments more closely resemble
the bundled NTFs in the Galactic center Radio Arc \citep{y-zmc84} than
the do the isolated NTFs.  

To summarize we have observed polarized nonthermal emission from several
linear structures with aspect ratios exceeding 10.  These
characteristics indicate that these filaments should be classified as
GC NTFs.

\subsection{The Spectral Index and Its Gradient}\label{sec:index}

\begin{figure}
\epsscale{0.65}
\rotatebox{-90}{\plotone{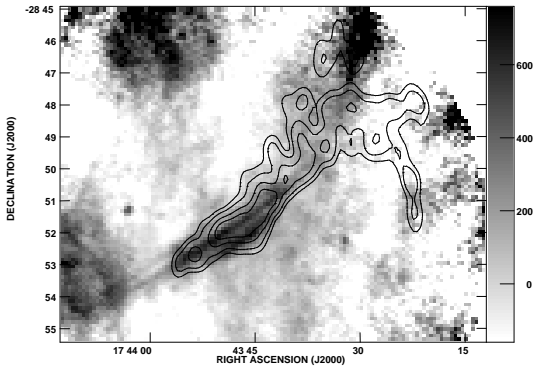}}
\caption[]{Grey scale of 6~cm emission (Figure~\ref{fig:6}) overlayed
on contours of the 90~cm emission (Figure~\ref{fig:90}).}
\label{fig:6vs90}
\end{figure}

The spectral index of \objectname[]{G359.85$+$0.39} must change along
its length because the peak in emission occurs at different locations
at the different wavelengths and because it is longer at 90~cm than
at~6 or~20~cm.  Figure~\ref{fig:6vs90} compares the 6 and~90~cm
emission.

The similar image resolutions at 90 and 20~cm make it is possible to
estimate the 20/90~cm spectral index with only a minimal convolution.
After convolving the 90~cm image to the 20~cm resolution, we made
cross-cuts perpendicular to the linear part and determined the peak
fluxes by fitting a two-point linear baseline at each cross-cut.  The
length of the cross-cuts was over~300\arcsec\ on each side of the
filament.  In almost all cross-cuts the endpoints were used to
establish the baseline.  However, it was also possible to choose
reasonable baselines using other points.  This leads to a range of
possible values for the peak flux at each cross-cut and hence a range
in spectral index at each position.

\begin{figure}
\epsscale{1}
\plotone{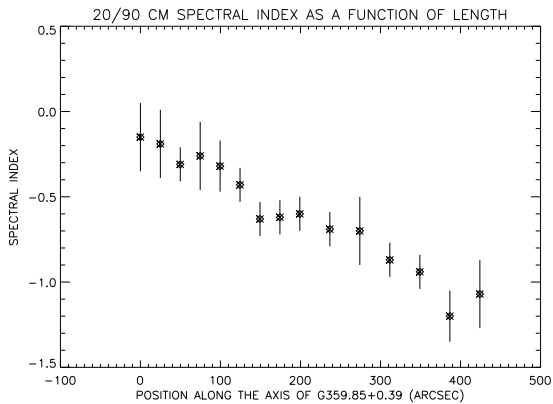}
\caption[]{The 20/90~cm spectral index as a function of length along
\protect\objectname[]{G359.85$+$0.39}.  The origin is at the southern
end of the filament.  Error bars were determine from basline
uncertainties in the individual cross-cuts.}
\label{fig:si}
\end{figure}

Figure~\ref{fig:si} shows the 20/90~cm spectral index plotted as a
function of length along \objectname[]{G359.85$+$0.39} with the
uncertainties on the points reflecting the uncertainties in the
baselines.  The spectral index varies smoothly from $\alpha = -0.15$
to~$-1.1$ from south to north, i.e., away from the \hbox{GC}.

There are two possible sources of systematic error in our
determination of the spectral index.  However, one source of
systematic error serves to \emph{flatten} the derived spectral
indices, i.e., the actual spectral index gradient may well be larger
than Figure~\ref{fig:si} shows.  First, the images from which the
spectral indices were determined do not have matching spatial
frequency coverage.  Even though the 20 and~90~cm images have similar
resolutions (see Figures~\ref{fig:90} and~\ref{fig:20sub}), the 90~cm
visibility data cover more of the inner $u$-$v$ plane than do the
20~cm visibility data.  We obtain a rough estimate of the effect of
the mismatch of spatial frequency coverage in the following manner:
The shortest antenna spacings in the VLA's DnC configuration are
approximately 35~m.  At~20~cm, these baseline lengths correspond to
angular scales of approximately 20\arcmin, meaning that structures on
scales of~10\arcmin--20\arcmin\ are poorly sampled or missing entirely
from Figure~\ref{fig:20} (and~\ref{fig:20sub}).  These angular scales
are comparable to or larger than the length of
\objectname[]{G359.85$+$0.39} at~20~cm, so we can approximate the 
effect of the missing short baselines as a constant offset to the
brightness of \objectname[]{G359.85$+$0.39}.  Thus, the spectral
indices shown in Figure~\ref{fig:si} may be too steep, but, if so,
the systematic bias should apply approximately equally to all.

The second possible systematic effect also occurs because of missing
short spacings, in particular the ``zero spacing,'' and is evident as
the large negative regions surrounding \objectname[]{Sgr~A} in
Figure~\ref{fig:20}.  Because \objectname[]{Sgr~A} is so bright, it
produces a large ``negative bowl'' surrounding it \citep[for an
additional illustration see][]{bv79}.  This negative bowl may extend
well past \objectname[]{G359.85$+$0.39}, but its gradient is directed
toward \objectname[]{Sgr~A}.  That is, the largest negative regions
occur near \objectname[]{Sgr~A}, and their magnitudes become
progressively smaller farther away from \objectname[]{Sgr~A}.  Thus,
this negative bowl would tend to depress the brightness at the south
end of \objectname[]{G359.85$+$0.39}, where the spectral index is
flattest, more so than at the north end.  We conclude that the
20/90~cm spectral index gradient shown in Figure~\ref{fig:si} is
robust and may actually be \emph{steeper} than shown.  Furthermore
there is no evidence of a negative bowl in the region of the
semi-circular emission.  Recall that the semi-circular emission would
not be detectable if the spectral index was steeper than -1.1.  Given
the spectral index is -1.1 in the linear structure and assuming the
gradient continues we would not expect to detect the semi-circular
emission.
 
We also attempted to use the same method to determine the 6/20~cm
spectral index along the length of \objectname[]{G359.85$+$0.39}.  The
differing spatial frequency coverage of the 6 and~20~cm observations is
far more problematic in this case.  The missing short spacings mean
that angular scales larger than about 25--35\% of the filament's length
are poorly sampled or missing.  Furthermore the wide difference in
resolution requires a large convolution.  We find that the 6/20~cm
spectral index is steeper than the 20/90~cm spectral index, with
$\alpha$ ranging from~$-0.9$ to~$-1.3$.   This result must be regarded
as an upper limit to the spectral curvature.  We also found no
systematic variation with position.  However, given the caveats
mentioned above we can not attach much significance to this result.
Additional observations are required to make definitive statements
about the 6/20~cm spectral index.

\section{A Spectral Index Gradient: Discussion}\label{sec:theory}

Previous studies of several NTFs \citep{lme99,lklh00} have found that
their spectral indices are constant with length.  For example, the
\objectname[]{Sgr~C} filament exhibits a constant 20/90~cm spectral
index $\alpha \approx -0.5$ over its entire length, about~27~pc.  One
prominent exception is the \objectname[]{Snake} filament, which shows
a spectral index gradient at the location of the major kink
\citep{gnec95}.  The index is steep at the kink with $\alpha \approx
-0.5$ and flattens to~0 moving away from the kink.

By contrast, \objectname[]{G359.85$+$0.39} exhibits a 20/90~cm
spectral index gradient, with the flattest spectral index ocurring at
the southeastern end of the filament, \emph{not} at the midpoint of
the filament where the flux peaks.  One might expect the location of
the peak flux to coincide with the acceleration site and the flattest
spectral index.  Clearly the electron acceleration in this object is
more complicated than this simple scenario.

In general, the coherence of the NTFs over many parsecs in a region
with a strong ram pressure from the surrounding molecular gas suggests
that the magnetic field in these structures is quite strong
\citep{y-zm87a,y-zm87b}.  Equating ram pressure with magnetic pressure
these authors estimate that the magnetic field of an NTF is of order
\hbox{1~mG}.  We adopt this value for \objectname[]{G359.85$+$0.39}.
The synchrotron emissivity corresponds to a number density of
relativistic particles of order $10^{-5}$~cm${}^{-3}$.  The background
particle density from observations of X-ray emitting gas in the GC
region is of order 1~cm${}^{-3}$ \citep{kmstty96}.  The electrons
responsible for the 90~cm emission have an energy of~0.14~GeV and a
synchrotron lifetime of $6 \times 10^4$~yr while the 20~cm emission is
from~0.29~GeV electrons with a lifetime of $2.9 \times 10^4$~yr.

As discussed above, although our estimate of the 6/20~cm spectral
index suggests spectral curvature between~6 and~20~cm, it is not
conclusive since we are missing 6~cm flux.  However, curved spectra
have been found by \cite{lme99} for both the Southern and Northern
NTFs and for the \objectname[]{Snake} filament by \cite{gnec95}.
\emph{We therefore argue that spectral curvature is a general property
of NTFs.}

Curvature is most commonly interpreted in terms of particle aging.
Our spectral index measurements indicate a break in the frequency
spectrum between~6 and~20~cm.  Assuming an initial power-law spectrum,
the time~$t$ for synchrotron losses to produce the observed curvature
is given by $t = 3.3 \times 10^4\,\mathrm{yr}\,B^{-3/2}\nu_b^{-1/2}$,
where the magnetic field~$B$ is in mG and the break frequency~$\nu_b$
is in GHz.  Using a magnetic field of~1~mG, a break frequency
between~6 and 20~cm corresponds to a source age of order $2 \times
10^4$~yr.  For comparison in a 1~mG field the synchrotron lifetime of
a 0.1~GeV electron emitting at a wavelength of~90~cm is of order $6
\times 10^4$~yr.  Although we cannot discount the possibility the
acceleration mechanism itself generated a curved spectrum and that
this is a young source, we will, for purposes of illustration, assume
an age of a few tens of thousands of years.

The origin and structure of the NTF magnetic fields is still a matter
of debate \citep[e.g.,][]{sl99}.  The central question is whether the
filaments are tracing a large-scale, globally-ordered magnetic field
or are independent local magnetic structures.  If part of an ordered
large-scale field, we would expect the NTFs are static equilibrium
magnetic structures.  Alternatively, if they are local, they are more
likely evolving dynamic structures.  In addition to the magnetic
field, the key questions surrounding the NTF phenomenon are the
location of the acceleration region and the subsequent transport
electrons.  There are no detailed models for the acceleration of
electrons in the NTFs, and it is often assumed that electron
acceleration occurs in a region considerably smaller than the size of
the filament \citep[e.g.,][]{sm94,rb96}.  The accelerated electrons
subsequently expand along the magnetic field and illuminate the rest
of the structure.  Particle transport is likely to be very different
in static structures than it is in dynamic ones.  We begin by showing
that this filament cannot be considered to be a static equilibrium
structure.  We then discuss the scenario we favor in which the
filament represents a localized enhancement in the GC magnetic field.

\subsection{Magnetic Field Structure and Particle
	Transport}\label{sec:transport}

We assume acceleration occurs locally and particles stream along the
(static) magnetic field. Such anistropic velocity distributions are
unstable and generate MHD waves which in turn scatter the particles.
Such resonant scattering \citep[e.g.,][]{w74,m82,dbm94} impedes
electron streaming and results in diffusive motion.  Resonant
scattering by low-frequency plasma waves produces electron pitch-angle
scattering without changing the energy of the electrons.  Pitch-angle
scattering randomly changes the direction of the electrons resulting
in diffusion parallel to the large-scale magnetic field.  For strong
pitch-angle scattering the electrons change direction rapidly and can
be confined to a small region of space \citep{dbm94}.  The magnitude
of the spatial diffusion coefficicient, $D_{xx}$ is inversely
proportional to the pitch angle diffusion coefficient $D_{\alpha
\alpha}$ as
\begin{equation}
D_{xx} = \frac{1}{6}\frac{V^2}{D_{\alpha\alpha}},
\label{eqn:diffuse}
\end{equation}
where $V$ is the particle velocity.  The pitch
angle diffusion coefficient for electron scattering by whistler or
Alfv{\'e}n waves (ignoring angular factors) is \citep{m82}
\begin{equation}
D_{\alpha\alpha}=\pi^2 e^2\frac{W(k_r)}{p\epsilon},
\label{eqn:adiffuse}
\end{equation}
where $k_r = eB/m_ec$ is the resonant wave number, $W(k_r)$ is the
wave energy density per unit wave number, $p$ is the particle momentum,
and $\epsilon$ the particle energy.  Expressing this coefficient in
terms of the background magnetic energy density $B^2/8\pi$ gives
\begin{equation}
D_{\alpha\alpha}
 \approx 4 \left(\frac{B}{1\,\mathrm{mG}}\right)\left(\frac{\epsilon}{1\,\mathrm{GeV}}\right)^{-1} \frac{k_rW(k_r)}{B^2/8\pi}.
\label{eqn:adiffuse2}
\end{equation}
For diffusive motion the root mean square displacement of a particle
is $\langle\Delta X^2\rangle = D_{xx}t$.  The parameter that
determines the transport is the ratio of the wave energy density to
the background magnetic energy density.  According to \cite{m82} the
maximum in the wave energy level occurs when the scattering reduces
the velocity anisotropy to the threshold anisotropy in one scattering
time.  For $(V_A/c) \sim 10^{-2}$, $k_r W(k_r)/(B^2/8\pi)$ is of order
$10^{-8}$ \citep [equation~7.76]{m82}.  We conclude that electrons
with an energy of 0.14~GeV diffuse about 13~pc in a magnetic field
of~1~mG in $4 \times 10^4$~yr.  Given that the lengths of
\objectname[]{G359.85$+$0.39} and other NTFs are of the same order of
magnitude, the age indicated by the spectral curvature is consistent
with this length if the acceleration occurs in a region whose length
is small compared to the length of the filament. However, such
diffusion acting alone does not produce a spectral gradient.

The parallel diffusion coefficient due to scattering by self-generated
waves depends on the first power of the energy.  Since the synchrotron
lifetime of an electron depends inversely on the first power of the
energy, the diffusive length associated with a given timescale is
independent of energy.  Spectral steepening could be produced by
diffusion that depends on a power of the energy that is less than 1.
For $D_{xx} \propto E^{\beta}$ with $\beta < 1$, then $\Delta X \propto
E^{-(1-\beta)/2}$.  For example, if $\beta =1/2$, then $\Delta
X\propto E^{-1/4}$ resulting in fewer higher energy particles at
large distances from the acceleration site.  A full mathematical
treatment, which is beyond the scope of this paper, is required to
establish if diffusion by pre-existing plasma turbulence could
generate the observed spectral steepening.  However, in a static
magnetic field configuration there is no compelling physical basis to
introduce a spectrum of turbulence. We therefore explore transport in
a dynamic configuration.

In a dynamic, evolving magnetic field configuration both plasma and
MHD turbulence are expected.  Furthermore the magnetic field in a
turbulent system could be tangled with a significant random component.
Polarimetric studies of other NTFs \citep[e.g.,][]{y-zwp97,lme99} show
that on the large-scales the magnetic field is aligned along the long
axis of the NTF suggesting an ordered structure.  However, on scales
much smaller than a beam (i.e., subparsec), the field could have a
significant random component.  Cross-field transport would dominate in
this situation.  Although classical theory predicts that the ratio of
parallel to cross field diffusion is of order $10^{-6}$, observations
of cosmic rays indicate a ratio of order $10^{-2}$ \citep{j99}.  A
reduction in our previous diffusion coefficient by this factor reduces
the diffusion length to~1~pc, which would certainly require
acceleration to occur along the entire length of the filament.
Electrons would clearly age before they could fill the filament, but it
is not at all clear the transport and distributed acceleration will
conspire to produce a uniform spectral gradient.  It seems in either
the static or dynamic configuration something more is required to
explain a uniform gradient.

\subsection{Magnetic Field Variation}\label{sec:bfield}

As an alternate possibility to explain spectral index variation, we
consider a combination of a curved particle spectrum and a spatially
varying magnetic field \citep{rk-sa92}.  In general particles of
energy~$E$ will emit synchrotron radiation most intensely at a
frequency $\nu_{\mathrm{ob}}=cB^2E$.  In a region of lower
magnetic field, higher energy particles will emit at a given
$\nu_{ob}$.  If the particle energy spectrum is curved, the result is
a gradient in the spectral index.  Assuming a smoothly declining
energy spectrum between two locations with magnetic field strengths of
$B_{1,2}$, the change in spectral index is $\delta\alpha
=a_c\log(B_1/B_2)$, where $a_c$ is the curvature of the frequency
spectrum \citep{br00}.  Unfortunately, in our case we have only three
measured frequencies, and it is not meaningful to fit a second order
polynomial to the spectrum and quantitately determine a curvature.

If a magnetic field gradient is responsible for the spectral gradient
in this source, the much larger northwestern extant of the source at
90~cm compared to~6 and~20~cm suggests a fairly rapid decrease in the
magnetic field at the this end of the filaments.  Such a rapid
divergence is \emph{not} consistent with a globally ordered magnetic field.
\cite{mm94,m96} suggests that the magnetic field in the GC region is,
to first order, a dipole centered on Sgr~A$^{\star}$ with a radius of
curvature comparable to the size of the central molecular zone, 100
to~200~pc.  The ratio $B/\nabla B$ is an estimate for the scale of
variation.  At a projected distance of~75~pc, this implies variation
on a scale of~25~pc.  However, the spectral index changes can be
measured on a scale of a few parsecs.  The region over which the
spectral index was measured is 13~pc.  Over this region 15 cross-cuts
were made, and changes in $\alpha$ occur over a few cross-cuts.  If a
magnetic field gradient is the cause for the spectral gradient, it
must be a local field.  We conclude that \objectname[]{G359.85$+$0.39}
is not tracing a large-scale magnetic field.
    
\section{Summary}\label{sec:conclude}

We have discovered a system of three parallel, nonthermal filaments
about~75~pc in projection from \hbox{Sgr~A}.  The key observational
feature of this system is a uniformly decreasing 20/90~cm spectral
index as a function of length.  It is also noteworthy that the peak
flux occurs about midway along the length of the filament, but the
flattest spectral index occurs at the end closest to the Galactic
center.

There are several explanations for the observed spectral gradient.
These include energy dependent diffusion in a turbulent plasma
magnetic field configuration and electron acceleration distributed
along the length of the filament.  However, independent of the
electron acceleration and transport, a simpler and more natural
scenario is that the magnetic field is varying along the length of the
filament.  A spectral gradient is a natural consequence of a curved
electron energy spectrum radiating in a decreasing magnetic field. If
so, the gradient scale length of a few parsecs is not consistent with
a large-scale field centered on \hbox{Sgr~A}.  This result suggests
that this nonthermal filament system is not necessarily tracing a
large-scale magnetic field.  Observations at additional frequencies
could be used to determine the spectral curvature and provide specific
constraints on the magnetic field structure.

\acknowledgements

We thank S.~Shore for several stimulating discussions, C.~Lang for
assistance with the observations and discussions in the early stages
of this work, and the referee for several suggestions that improved
the presentation of this work.  Basic research in radio astronomy at
the NRL is supported by the Office of Naval Research.  T.~N.~L.\ was
supported in part by a Master Scholarship grant from Kennesaw State
University.

\end{document}